\documentclass[sigconf,nonacm]{acmart}

\usepackage{soul}
\usepackage{color}

\usepackage{amsfonts}

\usepackage{amssymb}

\usepackage{amsmath}
\usepackage{algorithmic}
\usepackage{graphicx}
\usepackage{textcomp}

\usepackage{multirow}
\usepackage{stackengine}
\usepackage{listings}
\usepackage{url}

\definecolor{backcolour}{rgb}{0.95,0.95,0.92}
\definecolor{codegray}{rgb}{0.5,0.5,0.5}
\definecolor{commentcolor}{rgb}{0.5,0.5,0.5}
\newcommand{\toolname}{NVBitPERfi}

\AtBeginDocument{%
  \providecommand\BibTeX{{%
    \normalfont B\kern-0.5em{\scshape i\kern-0.25em b}\kern-0.8em\TeX}}}

\setcopyright{acmcopyright}
\acmYear{2023}
\setcopyright{rightsretained}
\acmConference[SC '23]{The International Conference for High Performance Computing, Networking, Storage and Analysis}{November 12--17, 2023}{Denver, CO, USA}

\acmDOI{10.1145/3581784.3607086}
\acmISBN{979-8-4007-0109-2/23/11}

\begin{document}

\title{Understanding the Effects of Permanent Faults in GPU's Parallelism Management and Control Units}


\author{Juan-David Guerrero-Balaguera}
\affiliation{%
  \institution{Politecnico di Torino -- Department  of Control and Computer Engineering}
   \city{Torino}
  \country{Italy}
}

\author{Josie E. Rodriguez Condia}
\affiliation{%
  \institution{Politecnico di Torino -- Department  of Control and Computer Engineering}
   \city{Torino}
  \country{Italy}
}

\author{Fernando F. dos Santos}
\affiliation{%
  \institution{Univ Rennes INRIA -- Taran Group}
   \city{Rennes}
  \country{France}
}

\author{Matteo Sonza Reorda}
\affiliation{%
  \institution{Politecnico di Torino -- Department  of Control and Computer Engineering}
   \city{Torino}
  \country{Italy}
}

\author{Paolo Rech}
\affiliation{%
  \institution{University of Trento -- Department of Industrial Engineering }
   \city{Trento}
  \country{Italy}
}

\begin{abstract}
    





Modern Graphics Processing Units (GPUs) demand life expectancy extended to many years, exposing the hardware to aging (i.e., permanent faults arising after the end-of-manufacturing test). Hence, techniques to assess permanent fault impacts in GPUs are strongly required, especially in safety-critical domains.

This paper presents a method to evaluate permanent faults in the GPU's scheduler and control units, together with the first figures to quantify these effects. We inject 5.83x10$^{5}$ permanent faults in the gate-level units of a GPU model. Then, we map the observed error categories as software errors by instrumenting 13 applications and two convolutional neural networks, injecting more than 1.65x10$^{5}$ permanent errors (1,000 errors per application), reducing evaluation times from several years to hundreds of hours. Our results highlight that faults in GPU parallelism management units impact software execution parameters. Moreover, errors in resource management or instructions codes hang the code, while 45\% of errors induce silent data corruption.

\end{abstract}






\received{\today{}}

\maketitle

\section{Introduction}
\label{sec_introduction}

GPUs are increasingly adopted in several fields, including High-Performance Computing (HPC), autonomous robots, automotive, and aerospace applications. The use of GPUs in applications that wander off their traditional fields (gaming, multimedia, and consumer market) has suddenly pushed the interest and posed questions, about their reliability~\cite{gpus_in_safety_critical_2018}.
 
 Currently, active GPU research targets the evaluation of reliability and the identification of feasible improvements. Most studies highlight a high sensitivity of GPUs to transient faults~\cite{tc2016, tr2019, luisEntrena2020Sensitivity, dueKojiro2021, Sulivan2021NVIDIADDR, Vallero2017, GPUQin, Nie2018, Yang21}, caused by the high number of available resources and the advanced semiconductor technology they adopt. Additionally, the parallelism management and control units of GPUs have been shown to be particularly critical since their corruption affects multiple threads~\cite{Rech_DSN2014, dueKojiro2021}. The parallelism of GPUs, which provides unquestionable benefit in terms of performance, is, then, one of the most vulnerable characteristics of the device. 
GPU manufacturers have provided effective reliability countermeasures by improving the memory cells design~\cite{selse2014}, adding error-correcting codes~\cite{ferraz_2022}, hardware structures for fault testing~\cite{nvidia_ist_2019}, and proposing software checksum~\cite{Siva22} or multi-threading redundancy~\cite{Wadden14}. 

Most of the available research on GPU's reliability targets \textit{transient} faults and their effects as software errors, leaving \textit{permanent} faults largely unexplored. This was justified since, in most applications 
the GPU's life expectancy does not exceed two years. However, GPUs employed in automotive, aerospace, and military applications are expected to be operative for many years. Additionally, typical operative conditions of HPC-grade GPUs, such as over-stress, high temperature, high frequency of operation, and technology node shrinking, are shown to accelerate aging~\cite{IRDS2022-MM} and even to expose the device to terrestrial radiation-induced permanent faults~\cite{IRPS_NASA_2021}. 

The extended utilization and premature aging suddenly raise questions about how GPUs and their applications behave in the presence of permanent faults. Crucially, only a few preliminary works target permanent faults in GPUs~\cite{nvbitfi2020, Haeseung18, guerrero_2022} and none focus on the parallelism management units.



In this paper, we aim to significantly advance the understanding of GPUs reliability by proposing a method to target a totally unexplored aspect: the effect of permanent faults in the GPU circuitry in charge of parallelism management. We decided to focus on the scheduler, the fetch, and decoder units since (a) they are the peculiar GPU resources mainly optimized for parallel operation, (b) a permanent fault affecting them will have non-trivial effects on the code execution, (c) they cannot be easily protected with error correcting codes or hardware redundancy, (d) they are likely to age faster than other units since they are always active during the execution of any software code. 

The challenge our methodology addresses is to provide an accurate, yet efficient, permanent faults impact evaluation. Injecting permanent faults directly in software would be fast~\cite{nvbitfi2020}, but is not a viable option for faults affecting the scheduler and control units, since a suitable fault model is missing. In fact, to accurately simulate the effect of permanent faults in the software, it is first necessary to track how \textit{each} machine instruction that uses the malfunctioning resource behaves. A detailed gate-level analysis is not a viable option either because of the extremely long simulation time. To characterize a simple CNN as LeNet on a gate-level GPU simulator would take more than 10,000 days! 

Inspired by other works in the field \cite{nimara_2016, Balasubramanian_2014, cho_2015, DOS_SANTOS_DNS_2021, sartor_2019}, we adopt a hybrid approach that combines accurate gate-level fault simulations with flexible software-level error injections. The behavior of a permanent fault is evaluated, at the gate level, using an open-source model of an NVIDIA GPU (FlexGripPlus~\cite{flexgripplus_2020}). The impact of the permanent faults on the execution of each machine instruction is evaluated and classified, identifying software error models to propagate in a real GPU. 
Then, to propagate these errors in software, we have crafted a dedicated software-based error injector framework (\toolname{}) able to automatically handle the error instrumentation in the kernel, to apply the corruption observed in the low-level evaluation, and to properly corrupt multiple threads and/or warps. 
With this approach, we characterize and discuss the reliability with respect to permanent faults of 15 real workloads, including two CNNs (LeNet and YOLOv3). 
The proposed frameworks, the gate-level analyses, the software-level reports, and the new \toolname{} tool are available in a public repository~\cite{REPO}.

\color{black}
    The main contributions of this work are:
    
    \begin{itemize}

    \item A method to identify the effects of permanent faults in terms of  errors at the instruction level; 
    
    \item The formalization of 13 categories of instruction errors \textit{(error models)} based on the effects of the permanent faults in the GPU's warp scheduler controller, fetch, and decoder unit;

    \item A new fault injector \textit{"\toolname{}"} built on top of NVBit, to map the 13 error models in software, instrument the code, and evaluate the permanent error effects in applications;

    \item An accurate understanding of why and how permanent faults in the GPU parallelism management units affect the execution of 15 real workloads.
    
    \end{itemize}

The remainder of the paper is structured as follows: Section~\ref{sec_background} provides the background and related work. Section ~\ref{sec_idea} describes the proposed multi-level fault injection evaluation. Section~\ref{low_experimental_results_Section} reports the gate-level fault injection results and identifies the error categories. Section~\ref{Software_based_erro_propagation} presents the software-level implementation of the error categories and evaluates their propagation effects in real applications. Section~\ref{sec_conclusion} concludes the paper. 

\section{Background and Related Work}
\label{sec_background}
     \begin{figure}[t]%
 	\centering
 	\includegraphics[width=0.48\textwidth]{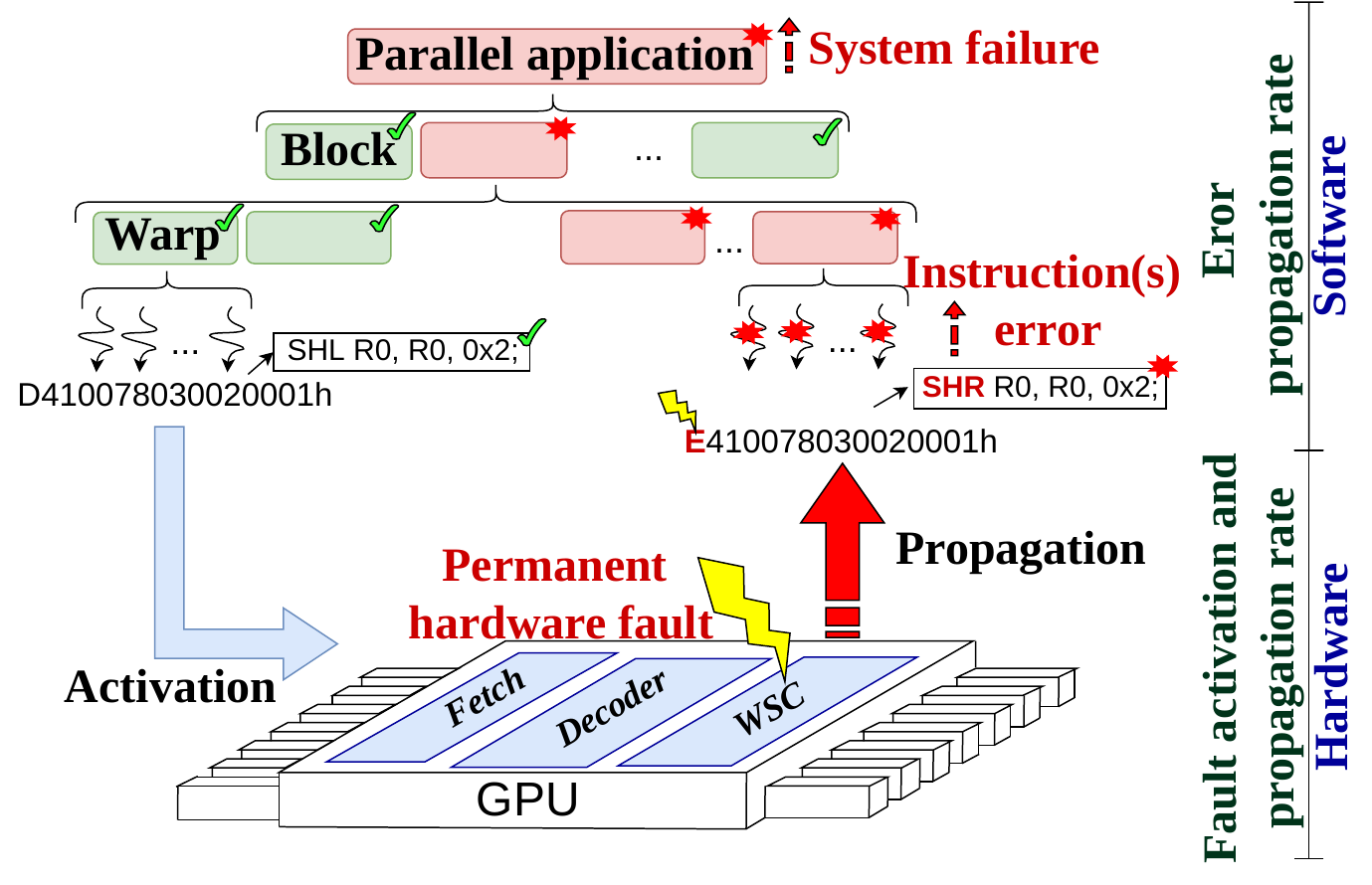}

\vspace{-2.5mm}
  
 	\caption{Propagation of permanent faults in GPUs. A \textit{fault} in the scheduler, fetch, or decoder unit produces nontrivial \textit{error} models on instructions and affects multiple threads/warps, causing a system \textit{failure}.}
  \par\vspace{-2.5mm}
 	\label{figure_example_errors}
     \end{figure}

In this section, we provide background and related works about GPUs reliability. Then, we discuss permanent faults and the challenges in their evaluation. Finally, we highlight the contributions and limitations of our study.

\subsection{GPUs architecture and reliability}


Modern GPUs 
are designed as arrays of parallel cores organized in processing clusters or \textit{Streaming Multiprocessors} (SMs). SMs {include} 2 to 4 \textit{Parallel Processing Block} (PPBs). One scheduler inside each SM statically distributes the tasks (thread-groups or \textit{Warps}) among the PPB cores. To manage parallelism and submit, distribute, and track warps into the available cores, each PPB includes a Warp Scheduler Controller (WSC), a fetch unit, and an instructions decoder unit~\cite{nickolls_2008, TIRUMALA_2017, cudaprogramming}. These are the units we characterize in this paper. 
%

Since GPUs have a large area with high availability of computing resources, they are particularly susceptible to experience hardware faults. 
A fault in the GPU hardware can have one of the following effects on the executed software. (1) \textbf{Masked}: no effect on the program output. The corrupted data is not used, or the circuit functionality is not affected. (2) \textbf{Silent Data Corruption (SDC)}: undetected output corruption, that is, the application finishes, but the output is not correct. (3) \textbf{Detected Unrecoverable Error (DUE)}: program or system crash.
The GPU reliability to \textit{transient} faults has already been evaluated through microarchitectural and low-level fault simulation~\cite{gpuFI4, flexgripplus_2020, fgpu_2016}, software-based fault injection~\cite{LLFI2014, Vallero2017, GPUQin, Nie2018, Yang21}, and beam experiments~\cite{tc2016, tr2019, luisEntrena2020Sensitivity, dueKojiro2021, Sulivan2021NVIDIADDR}. Since GPUs execute several processes in parallel, it has been shown that a transient corruption in the WSC or a single error in shared resources affects various output elements~\cite{Rech_DSN2014, tc2016, luisEntrena2020Sensitivity, Sulivan2021NVIDIADDR, Siva22}.


\subsection{Permanent faults on GPUs}

\label{Permanent_faults_definition}
\label{EPR_definitions}

Permanent faults represent a new severe risk for GPUs' reliability. In fact, while GPUs used in multimedia are changed every few years, GPUs employed in various domains, such as the automotive, military, and aerospace ones, have an extended life expectancy. Additionally, the \textit{"International Roadmap for Devices and Systems - 2022"} (IRDS) and independent studies~\cite{hamd_2013, strojwas_2019} state that modern digital devices implemented with cutting-edge technologies (beyond 7nm) are highly susceptible to \textit{Electromigration} and \textit{Time-Dependent-Dielectric-Breakdown}, both major sources of accelerated aging and permanent faults \cite{IRDS2022-MM,Blackeq}. Crucially, IRDS emphasizes that the device's lifetime decreases by half at each new manufacturing process generation\cite{IRDS2022-MM}. This, in combination with over-stress operative conditions (typical for GPUs), high-temperature, and radiation-induced latch-ups, exacerbate the probability of having permanent faults \cite{JaeGyungProdLifetime7nm,IRPS_NASA_2021, Constantinescu_2003, pae_2008, prasad_2013}. 

    A permanent \textit{fault} is a physical defect in a circuit, unit, or device. When the applied stimuli activate the permanent fault, it can propagate to a visible software state, corrupting data or operation output, thus becoming an \textit{error}. If the error propagates and affects the outputs produced by the system, it becomes a \textit{failure} causing a crash, hang, or silent data corruption. 
    As shown in Figure~\ref{figure_example_errors}, we track permanent faults propagation from the hardware to the software output using two metrics: the \textit{Fault Activation and Propagation Rate} (FAPR), that measures the probability of a permanent hardware fault to be activated by stimuli (e.g., program's instructions) and then propagate reaching a software visible state (thus becoming an error). Then, the \textit{Error Propagation Rate} (EPR) measures the probability for errors (caused by a permanent fault) to propagate till the output, becoming a failure.



    
Several works performed extensive analyses of possible sources of permanent faults in processor-based systems~\cite{henkel_2013, amrouch_2015}. Other studies wisely focused on identifying error models at higher levels to simplify the analysis \cite{fang_to_remove_2016}. In \cite{gizopoulos_2021}, the authors emphasize the importance of fine-grain low-level and cross-layer resilience evaluations, highlighting the weakness of purely software error propagation. Inspired by these insights, we combine low-level and software-level fault injections, to achieve high-reliability evaluation accuracy. 

Only a few preliminary studies evaluate the incidence of permanent fault effects  in GPUs. In~\cite{defour_2013}, the authors investigate the effect of permanent faults in GPUs by increasing the temperature and accelerating the aging process. Other works evaluate GPU memory permanent faults affecting the weights of a CNN~\cite{lofti_2019}. 
Some permanent fault injectors have also been proposed. In \cite{guerrero_2022}, the authors proposed a customized software-based error injector to evaluate the effect of permanent faults on the register file and functional units. Additionally, NVBitFI~\cite{nvbitfi2020} allows the user to inject permanent faults. Unfortunately, in all the available permanent fault injectors, the proposed error models consider only limited hardware units (mostly memory resources and functional units).

None of the studies on GPUs have focused their evaluations on the parallelism management units, as we do in this paper. Previous works tools can hardly be extended to evaluate more complex and critical units, such as the parallelism management units. 

\subsection{Contributions and potential limitations}

This is the first paper proposing a methodology and showing results about the propagation of permanent faults in the GPU parallel management and control units (i.e., WSC, decoder, and fetch units).
The criticality of these units and the possible impact of their failures in the threads/warps execution require a dedicated and accurate study. The complexity of the evaluation we propose is totally different from previous works that focus on GPU memory or functional units failures~\cite{nvbitfi2020, guerrero_2022}. In fact, besides the effect on the instruction execution, depending on the fetched/decoded/scheduled instruction, a permanent corruption in the GPU parallel management units can modify (a) the opcode or operands, (b) the control-flow of the code, (c) the threads/warps status (enable/disable), (d) the assignation or enabling of a GPU resource. Additionally, as depicted in Figure~\ref{figure_example_errors}, all these corruptions can affect one or multiple threads or warps.

We adopt a multi-level fault injection approach, separating the accurate low-level simulation from the software-level propagation. This methodology has been shown effective in reducing the simulation time in other works in the field \cite{condia_vts_2021, DOS_SANTOS_DNS_2021, nimara_2016, cho_2015}, but has never been applied to GPU's parallel management units. Thanks to our multi-level analysis, we can track permanent faults propagation in real-world applications.

The proposed methodology, thanks to the documentation included in the public repository~\cite{REPO}, can be extended to other units in the GPU, to other devices, and to other fault models (delay, intermittent, or transient faults). 
Despite the generality of our methodology, we acknowledge that our evaluation has some intrinsic limitations. In fact, the low-level characterization uses one of the few available open-source GPUs (FlexGripPlus). The implemented GPU's Instruction Set Architecture (ISA) in the available model is old (G80), which is a common constraint for research works targeting commercial devices~\cite{BRAGA_2021, condia_vts_2021}. Thus, the distribution of error models might be biased by the specific GPU implementation. However, since FlexGripPlus is CUDA compliant, the behavior of the faults is expected to be similar also in modern GPUs, and, as mentioned, our methodology can be adapted to more advanced models as they become available. Furthermore, since our method is focused on units inside SMs, novel GPU features (such as concurrent streams and multi-kernels, that act at a higher architectural level) will not undermine the accuracy of our evaluation. In fact, modern features are handled by task schedulers or additional hardware that then end up assigning the blocks to the internal units in the SM that we are evaluating.

\color{black}
\section{Proposed methodology}

    \begin{figure*}[t]%
	\centering
	\includegraphics[width=0.92\textwidth]{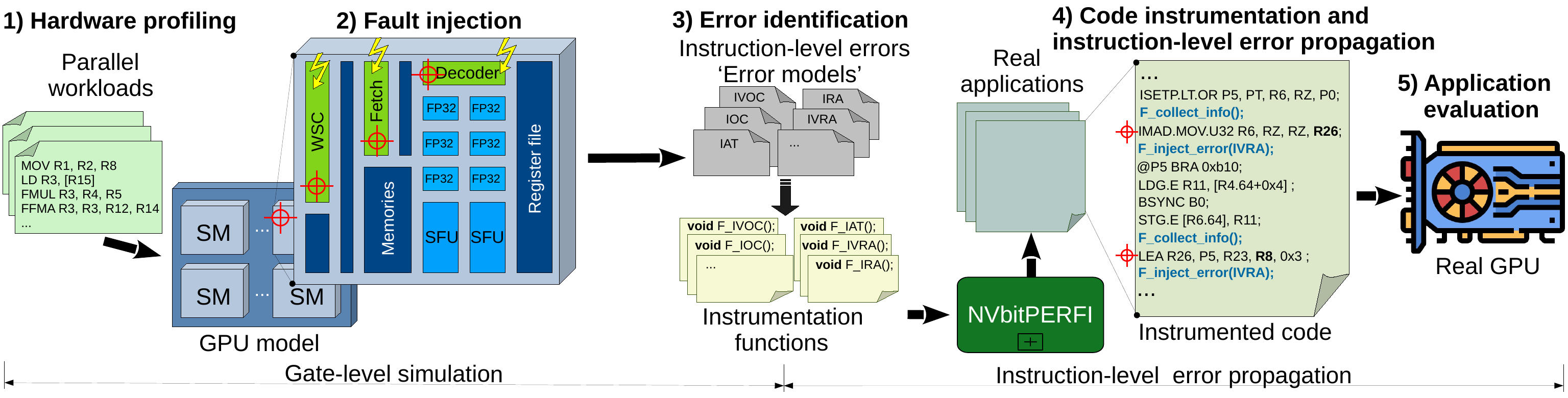}

\vspace{-2.5mm}
 
	\caption{A general scheme of the method to characterize fault effects in parallelism management units of GPUs.}
	\label{fig_general_scheme_proposed_method}
 \par\vspace{-1\baselineskip}
    \end{figure*}


\label{sec_idea}
    
    
    
    This section describes the main idea and the steps of the proposed method, see Figure~\ref{fig_general_scheme_proposed_method}. The \textit{low-level} evaluation exploits the accuracy of gate-level simulation to classify fault effects in terms of instruction errors. The \textit{high-level} part employs a time-efficient software-based fault injector on real GPUs to assess the effect of permanent faults on complete applications. The method comprises five steps: \textbf{1.} hardware unit profiling, \textbf{2.} gate-level fault injection, \textbf{3.} error identification, and classification, \textbf{4.} code instrumentation and instruction-level error propagation, and \textbf{5.} application evaluation and failure classification. 
     The next subsections detail each step. 
    
    \subsection{Hardware Unit Profiling}

    
    In this step, the unit profiling resorts to the characterization of each instruction from several representative parallel workloads. In particular, every dynamic instruction is executed on the GPU and we collect the accurate golden (fault-free) operation from the targeted hardware units (\textit{WSC}, \textit{fetch}, and \textit{decoder}). We resort to the gate-level netlist of the unit to test, while the rest of the GPU is simulated at the Register Transfer Level (RTL). The GPU mixed implementation (gate-level for the units of interest, RTL for the rest) allows to collect and trace per-cycle information on the tested unit at the gate level and keep the interaction with the other units at RTL. This step provides the golden copy of all the unit signals, including the input patterns.
    
    We developed a \textit{profiling hardware mechanism} tool to instrument the GPU model and profile the hardware unit utilization. For each instruction, our tool collects structural and operational information from the unit, including \textit{(i)} the status of primary inputs and outputs, \textit{(ii)} the timing information for the instruction, \textit{(iii)} the instruction's type, and \textit{(iv)} the time intervals of each performed operation. 
    
    
    


    \subsection{Gate-Level Fault Injection}


    This step characterizes permanent \textit{stuck-at} faults on each possible fault site in a targeted unit. Exhaustive gate-level fault injection campaigns are essential to determine if a fault is activated, propagated, and how it possibly manifests as an error at the output of the evaluated component. The simulation complexity would explode if we consider all possible stimuli combinations for each fault characterization. 
    Therefore, as mentioned in the previous step, we select as stimuli the individual instructions \textit{(patterns)} extracted from representative workloads. 
        

    The effect of a permanent fault can manifest at any point in time, depending on the instructions (\textit{stimuli}). Thus, we exhaustively evaluate the individual execution of every instruction and the activation and propagation of each individual fault. 
    To track the propagation of any possible effect on the outputs of the unit, we verify the fault propagation after the execution of the instruction by comparing the golden instruction profile with the current faulty operation. Then, we collect the observed effects and the exciting pattern (instruction information), which later serve to identify and correlate the fault model with the corrupted instruction. It is worth noting that we track the execution of the complete instruction across the GPU architecture to guarantee the identification of any possible fault propagation, so allowing the characterization of masking, hanging, and latent (inactive) effects.
    
    




    
    
    \subsection{Error Identification and Classification}
    
    \label{error_ident}


    In this step, we correlate the hardware profiling and the fault injection results to identify hardware fault effects in terms of visible instruction errors (e.g., change of operand or incorrect addressing to memory), hence modeled in software. We build a list of possible instruction errors caused by the injected permanent faults. The error models from the low-level fault injections are classified according to the corrupted functionality and effects on the software's visible state of every instruction. Thus, the error models represent the mapping of hardware faults as instruction errors. Given the parallel architecture of GPUs, a fault might corrupt the instruction execution in one or multiple threads, and in one or multiple warps. We identify and classify errors considering both cases.

    \subsection{Instruction-level Error Propagation}
\label{Code_instrumentation_methodology}

    \begin{table}[tb]
\centering
\caption{Codes used for the software-level error injections.}
\label{tab:apptable}

\vspace{-2.5mm}
\begin{tabular}{cccc}
\hline
\textbf{}          & \textbf{Data type} & \textbf{Domain}     & \textbf{Suite} \\ \hline
\textbf{vectoradd} & FP32               & Linear algebra      & CUDA SDK       \\
\textbf{lava}      & FP32               & N-body              & Rodinia        \\
\textbf{mxm}       & FP32               & Linear algebra      & CUDA SDK       \\
\textbf{gemm}      & FP32               & Linear algebra      & CUDA SDK       \\
\textbf{hotspot}   & FP32               & Structured Grid     & Rodinia        \\
\textbf{gaussian}  & FP32               & Linear algebra      & Rodinia        \\
\textbf{bfs}       & INT32              & Graphs              & Rodinia        \\
\textbf{lud}       & FP32               & Linear algebra      & Rodinia        \\
\textbf{accl}      & INT32              & Graphs              & NUPAR          \\
\textbf{nw}        & INT32              & Dyn. Programming & Rodinia        \\
\textbf{cfd}       & FP32               & Unstructured Grid   & Rodinia        \\
\textbf{quicksort} & INT32              & Sorting             & CUDA SDK       \\
\textbf{mergesort} & INT32              & Sorting             & CUDA SDK       \\
\textbf{lenet}     & FP32               & Deep Learning       & Darknet        \\
\textbf{yolov3}    & FP32               & Deep Learning       & Darknet        \\ \hline
\end{tabular}
\vspace{-3mm}
\end{table}

  Once the permanent hardware fault effect has been characterized as a visible software effect, we can propagate it through representative applications using fast software-based error injection in real GPUs. To do so, we implemented one software \textit{error function} for each permanent error model obtained in the error identification and classification experiment (step 3). These error functions are inserted in the application's SASS code to implement the permanent error effect during the application's execution, mimicking in software the equivalent effect of a permanent fault in the GPU unit.
  
  
  We crafted a customized binary instrumentation tool (\textit{\toolname{}} \cite{REPO}) built over the NVBit framework \cite{nvbit2019} to propagate the instruction-level error through the application software. To implement the software-level error propagation at the ISA-level, our tool adopts the Hardware-Injection through Program Transformation (HIPT) technique~\cite{nvbitfi2020}. \toolname{} mimics the behavior of a permanent error during the application's execution considering: \textit{i)} the error model specifications, \textit{ii)} the GPU architecture details, and \textit{iii)} the instrumentation mechanisms offered by NVBit.

    \begin{table}[]
\centering
\caption{Tested units area and utilization percentage w.r.t. a FP32 functional unit.}
\label{tab:table_area_units}
\vspace{-2.5mm}
\begin{tabular}{cccc}
\hline
\textbf{Unit}               & \textbf{Area ($nm^2$)} & \textbf{\begin{tabular}[c]{@{}c@{}} FP32 core (\%)\end{tabular}} & \textbf{\begin{tabular}[c]{@{}c@{}}Utilization (\%)\end{tabular}} \\ \hline
\textit{\textbf{WSC}}       & 11,854.4               & 114.3                                                                            & 100.0                                                               \\
\textit{\textbf{Decoder}}   & 760.8                  & 7.3                                                                              & 100.0                                                               \\
\textit{\textbf{Fetch}}     & 708.2                  & 6.8                                                                              & 100.0                                                               \\
\textit{\textbf{FP32 unit}} & 10,367.8               & 100.0                                                                            & $\sim$(10.0 - 40.0)                                                 \\ \hline
\end{tabular}
\vspace{-3mm}
\end{table}





We addressed two main challenges in the implementation of \toolname{}. (1) The target hardware units corruption can impact one or multiple threads in one or multiple warps. (2) Since we are addressing permanent faults, each instruction mapped to the corrupted hardware unit must be corrupted. Thus, we need to identify all the instructions activating the fault. 

It is crucial to have detailed hardware error specifications to identify how many threads/warps need to be affected by the permanent hardware fault. 
An error, for instance, may disable/enable one or multiple threads by interchanging their execution with a set of threads from the same warp or different warps (e.g., $<Thread_0$,$Warp_0>$ issues the $<Thread_{17}$,$Warp_8>$, and this produces the skipping execution of the $<Thread_0$ $Warp_0>$). 
To identify the instructions mapped to the corrupted hardware, we consider GPU's architectural details, that denote the parallelism specifications, such as the maximum number of resident warps/threads per Streaming Multiprocessor (SM), and the number of sub-partitions that every SM contains. We use these architectural functionalities to define an error descriptor that links the physical defects of hardware units under analysis and the portions of the parallel application where the error will take effect. We consider the following fields: \textit{i)} the SM identifier number, \textit{ii)} the sub-partition identifier (PPB), \textit{iii)} the set of warps associated with the sub-partition, \textit{iv)} the target threads inside the selected warps, and \textit{v)} additional parameters related to the specific error model, such as targeted operands, opcodes, error bit-masks, etc.

\subsection{Applications Evaluation}

Once the permanent fault effect has been characterized as software error models and the procedures to corrupt thread(s)/warps(s) in a real GPU have been implemented, we can effectively evaluate the impact of permanent faults on real workloads by using the \textit{\toolname{}} framework implementing the HIPT technique. This methodology reduces the simulation times by several orders of magnitude compared to the classical logic simulation approach. For example, an entire fault injection campaign for all the error models using our methodology for the GEMM code can be performed in less than 24h, while using only low-level fault injections, the same campaign would take 60,000 hours (i.e., more than 6 \textit{years}). 

We select 15 realistic workloads (listed in Table~\ref{tab:apptable}) to evaluate the error models implemented on \toolname{}. To demonstrate how the methodology can be applied to any application, we select workloads from various domains, including Deep Learning, Linear algebra, N-body simulation, and Graphs. 
The selected codes are instrumented and executed inside \toolname{} and the permanent fault outcome is characterized as masked, SDC, or DUE. 

    
    

\section{Low-level Fault Characterization}

    \label{low_experimental_results_Section}
        
   
    
    In this Section, we present the results of the gate-level permanent fault injection experiments performed in GPU parallelism management units using the FlexGripPlus GPU model. We configure FlexGripPlus with one PPB per SM cluster, and 32 SP cores per PPB. The gate-level implementations of the units are obtained using a 15nms Open Cell Library \cite{martins_2015_15nms}. Table \ref{tab:table_area_units} shows the percentage of area occupied by each unit, compared to one FP32 functional unit core, and their utilization percentage, taken from profiling several workloads (described below). Despite the relatively low area of the fetch and decoder units, these units are of paramount importance in the execution of instructions since they are continuously stimulated by every instruction (while the FP32 unit is stressed, on average, only by 10\% to 40\% of instructions), thus accelerating aging. Despite the relatively small area of the units we target, their continuous operation and their failure criticality motivate our study.

    The low-level evaluation starts with the golden unit hardware profiling of the \textit{WSC}, \textit{fetch}, and \textit{decoder}. In this case, we identify the signals of interest and the golden (fault-free) unit outputs. We use \textit{all} the dynamic instructions (more than 25,200 in the real code) from 14 representative parallel workloads from Rodinia and NVIDIA SDK benchmarks (\textit{Sort}, \textit{Vector\_Add}, \textit{FFT}, \textit{Tiled Matrix Multiplication}, \textit{Na\"ive Matrix Multiplication}, \textit{Reduction}, \textit{$Gray\_Filter$}, \textit{Sobel}, \textit{Scalar Vector Multiply}, \textit{Nn}, \textit{$Scan\_3D$}, \textit{Transpose}, \textit{$Euler\_3D$}, and \textit{Back Propagation}). 
    Then, the fault evaluation resorts to 42 localized fault injection campaigns (one per benchmark for each of the three units) on an industrial-grade logic simulator (\textit{ZOIX} by \textit{Synopsis}) to evaluate the execution of every individual dynamic instruction from the workloads (i.e., the equivalent \textit{exciting pattern} activating a unit) and identify the fault propagation effects. This procedure evaluates 708,808 permanent faults (i.e., the whole stuck-at-fault list) from the WSC (426,092), fetch (130,480), and decoder (152,236) units, respectively. The hardware profiling and the fault injection campaigns are performed on a server machine which includes 12 Intel Xeon CPUs running at 2.5 GHz and with 256 GB of RAM. It is worth noting that extensive multi-threading schemes (from 10 up to 40 parallel processes) are used to speed up the fault evaluation campaigns.
    


    Table~\ref{tab:table_HW_fault_rate} first reports the total number of considered stuck-at faults for each unit and classifies faults in the following categories:

    \begin{itemize}
    
        \item \textit{Uncontrollable} faults (125,808), i.e., those  permanent faults that are never activated or propagated by any input stimuli.

        \item \textit{Hardware Masked} faults, i.e., faults that are activated by the input stimuli but whose effect never reaches the unit outputs (30.0\% in the WSC, 24.5\% in the fetch, and 22.2\% in the decoder) in any of the executed instructions. These faults are thus innocuous and can be discarded from our analysis.
    
        \item Permanent faults that cause a  \textit{hardware hang}, so the GPU stops responding or unit's ports are corrupted, e.g., high-impedance. Only 1.2\% to 3.5\% of the faults caused a hang. A detailed analysis shows that most hang sources handle control signals (e.g., state machine control signals) or synchronization signals among the units (e.g., pipeline).

        \item \textit{Software errors:} faults that reach one or more unit's outputs and can corrupt the software. These faults are highly likely, being 30.5\% of injections for the \textit{WSC}, 47.39\% for the \textit{fetch}, and 49.29\% for the \textit{decoder} unit. These faults corrupt the unit's outputs handling or selecting instruction's parameters, such as the memory type or the thread(s)/warps(s) status.
    \end{itemize}
  
     \begin{table}[t]
\centering
\caption{Percentage of faults that are uncontrollable, masked, cause hangs or instruction-level errors.}

\vspace{-2.5mm}

\label{tab:table_HW_fault_rate}
\begin{tabular}{cccccc}
\hline
\textbf{Unit}    & \textbf{Total} & \textbf{\begin{tabular}[c]{@{}c@{}}Uncontrol-\\ lable\end{tabular}} & \textbf{\begin{tabular}[c]{@{}c@{}}HW\\ Masked\end{tabular}} & \textbf{\begin{tabular}[c]{@{}c@{}}HW\\ Hang\end{tabular}} & \textbf{\begin{tabular}[c]{@{}c@{}}SW\\ errors\end{tabular}} \\ \hline
\textbf{WSC}     & 29,850         & 35.9\%                                                              & 30.0\%                                                       & 3.6\%                                                      & 30.5\%                                                       \\
\textbf{Fetch}   & 9,320          & 26.9\%                                                              & 24.5\%                                                       & 1.2\%                                                      & 47.4\%                                                       \\
\textbf{Decoder} & 10,874         & 26.0\%                                                              & 22.2\%                                                       & 2.5\%                                                      & 49.3\%                                                       \\ \hline
\end{tabular}
\vspace{-3mm}
\end{table}

    To further categorize the faults in the last category, i.e., those that produce instruction-level errors on any instruction from the real code, we analyze the hardware profiles, the fault injection campaign results, and the structural information of the GPU.
    We have identified four main error groups ({\textit{i)}} operation, {\textit{ii)}} control-flow, {\textit{iii)}} parallel management, and {\textit{iv)}} resource management errors), which are further  divided into 13 types of errors affecting any software instruction, as follows.
    
    \subsubsection{Operation errors}

        \begin{itemize}
    
        \item \textbf{Incorrect Operation Code Error (IOC):} the operational code of an instruction is modified and still valid, but the executed instruction type (or its parameters) is different.
        
        \item \textbf{Invalid Operation Code Error (IVOC):} the opcode of the instruction is modified and not valid.
        
        \item \textbf{Incorrect Register Addressed Error (IRA):} an incorrect (yet valid) register is addressed, affecting the instruction.
        
         \item \textbf{Invalid Register Addressed Error (IVRA):} an incorrect and not valid register is addressed (i.e., a register outside the limit of registers per thread).
        
        \item \textbf{Incorrect Immediate Operand Error (IIO):} the immediate operand is corrupted.
    \end{itemize}
    
    \subsubsection{Control-flow errors}
    
    \begin{itemize}
    
        \item \textbf{Work-flow Violation Error (WV):} the workflow of an instruction is modified by corrupting the predicate conditions.
    \end{itemize}

    \subsubsection{Parallel management errors}

    \begin{itemize}
    
        \item \textbf{Incorrect Parallel Parameter Error (IPP):} incorrect addressing of resources shared among the warp, such as the shared memory and register files regions.

        \item \textbf{Incorrect Active Thread Error (IAT):} unauthorized enable or disable of threads in a warp.

        \item \textbf{Incorrect Active Warp Error (IAW):} incorrect detention, assignation, or unauthorized submission of a warp.

        \item \textbf{Incorrect Active CTA Error (IAC):} incorrect detention, assignation, or unauthorized submission of a CTA (cooperative thread array) in the GPU core.
    \end{itemize}
    
    \subsubsection{Resource management errors}

    \begin{itemize}
    
        \item \textbf{Incorrect Active Lane Error (IAL):} unauthorized enable or disable of lanes in a GPU core.
        
        \item \textbf{Incorrect Memory Source Error (IMS):} incorrect assignation of a memory resource for operand loading.

        \item \textbf{Incorrect Memory Destination Error (IMD):} incorrect assignation of a memory resource for result's storing.

    \end{itemize}


    We differentiate errors that cause an \textit{incorrect} {operand} from those that cause an \textit{invalid} operation or action. While both types of errors modify the same instruction field (e.g., both IOC and IVOC modify the opcode), the former is likely to induce a data error since a (wrong) instruction is executed or a (wrong) memory value is read/written, while the latter blocks the execution. It is worth noting that most errors affect the thread management units and the parallelism in the GPU. Thus, most error types (IOC, IVOC, IRA, IVRA,  IPP, IAW) affect all threads in a warp, while others (IIO, WV, IAT, IAC, IMS, and IMD)  mainly corrupt one or a few threads per warp. The information about multiple threads/warps corruption is used in Section~\ref{error_model_implementation} to map the effects into instruction-level errors.

    \begin{figure}[t]%
	\centering
	\includegraphics[width=0.48\textwidth]{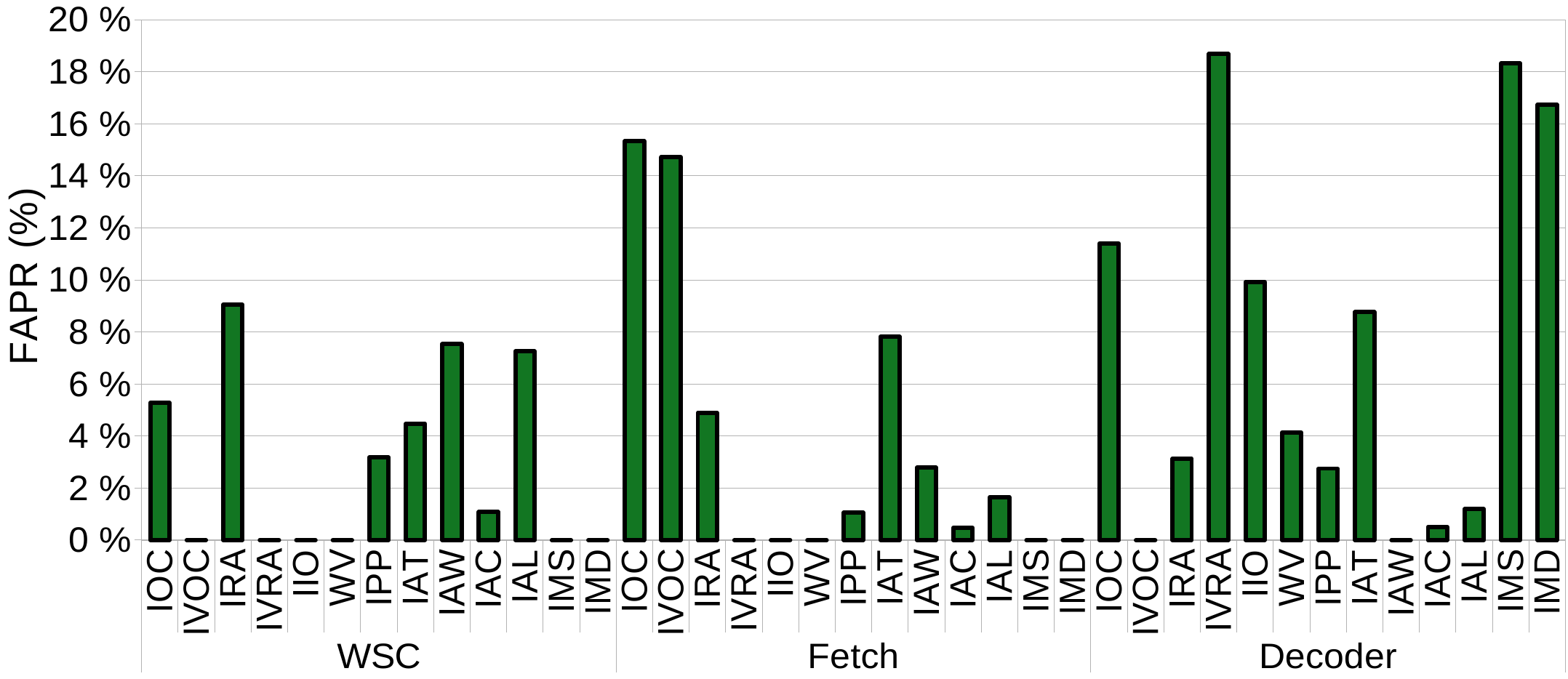}
    \par\vspace{-1\baselineskip}
	\caption{Fault Activation and Propagation Rate (FAPR) 
    for the identified faults as SW errors in the WSC, fetch, and decoder units. Faults are grouped by error types.}
	\label{AVF_figure_micro_arch}
    \par\vspace{-1.5\baselineskip}
    \end{figure}

 
 Figure~\ref{AVF_figure_micro_arch} shows the FAPR, i.e., the probability for a hardware permanent fault to be activated and to propagate to a software visible state. We separate the FAPR of permanent faults injected in each of the three units to cause one or more of the identified error types. The most common instruction error models are IVRA, IMS, and IMD in the decoder unit, IVOC in the fetch unit, and IOC for all units. On the contrary, some instruction error classes are highly unlikely to occur (e.g., from 0.48\% of IAC in WSC, to 7.53\% of IAW in the fetch unit). The low percentage of IAC errors (i.e., wrong block scheduling) is explained by noting that the considered WSC, the fetch, and the decoder units handle finer-grain parallel management (operation of threads and warps), instead of coarse-grain (CTAs) parallel management. Interestingly, faults in the decoder unit cause a wider spectrum of possible instruction effects (11 out of 13 error categories). This is  due to the fact that the decoder directly interacts with the machine code of the instructions.


    
We also identified some single permanent faults causing more than one error type. This is unsurprising since we are considering permanent faults that can be activated differently based on the stimuli. We have seen that (a) the same permanent fault may produce different types of software errors (from 1.28\% to 14.9\% for the WSC, about 1.98\% in the fetch, and less than 0.25\% in the decoder unit, depending on the executed instruction), and (b) the same permanent fault may simultaneously produce two or more types of software errors during the operation of a single instruction (less than 18.4\% of faults). Since we keep track of the instruction opcode and input stimuli that activated the permanent fault, we can correlate the error model to inject in software (Section~\ref{sw_experimental_results}) with the instruction being executed. This information allows to propagate the hardware fault in software and to understand the probability for a  permanent fault to be activated in a realistic application.

\section{Software-based Permanent Error Propagation}

\label{Software_based_erro_propagation}
This section describes the environment we developed to analyze the propagation of errors at the software level and discusses the results. 

\begin{figure}
    \centering
    \footnotesize 
    \begin{minipage}{0.47\textwidth}
    \begin{lstlisting}[title={a) Destination operand field case},abovecaptionskip={-2pt}]  
    (* $\cdots$ *) 
    /** Error function: Part I*/
    (*$M\footnotemark[1] \Leftarrow Rd[T_x,W_x]\footnotemark[2]$*)
    /**Target SASS instruction*/
    (*\textbf{IMAD}*) Rd, Rsx, Rsy, Rsz
    /** Error function: Part II*/
    (*$R_{IR}\footnotemark[5] \leftarrow Rd\footnotemark[4] \oplus bitErrMask\footnotemark[3]$*)
    (*$R_{IR}[T_x,W_x] \Leftarrow Rd[T_x,W_x]$*)
    (*$Rd[T_x,W_x] \Leftarrow M$*)
    (* $\cdots$ *) \end{lstlisting}
    \end{minipage}\par\vspace{-1\baselineskip}
    \begin{minipage}{0.47\textwidth}
    \begin{lstlisting}[title={b) Source operand field case},abovecaptionskip={-2pt}]  
    (* $\cdots$ *) 
    /** Error function: Part I*/
    (*$R_{IR} \leftarrow R<sx,sy,sz>\footnotemark[6] \oplus bitErrMask$*)
    (*$M \Leftarrow R<sx,sy,sz>[T_x,W_x]$*)
    (*$R<sx,sy,sz>[T_x,W_x] \Leftarrow R_{IR[T_x,W_x]}$*)
    /**Target SASS instruction*/
    (*\textbf{IMAD}*) Rd, Rsx, Rsy, Rsz
    /** Error function: Part II*/
    (*$R<sx,sy,sz>[T_x,W_x] \Leftarrow M$*)
    (* $\cdots$ *) \end{lstlisting}
    \end{minipage}
\vspace{-4.5mm}
    \caption{Description of IRA/IVRA error models.}
    \label{fig:IRAerr}
    \par\vspace{-0.5\baselineskip}
\end{figure}

\subsection{Error model implementation/propagation}
\label{error_model_implementation}
We implement, as software procedures, the permanent error models derived from the fine-grain circuit-level analysis in \toolname{} to mimic in detail each error according to their specifications. 



We use the same philosophy of NVBitFI: we devised dedicated instrumentation functions 
inserted in the assembly source code of the GPU kernels during the instrumentation stage \cite{nvbit2019}. Then, the error is injected, propagated and evaluated (at speed) once the \textit{faulty kernel} is issued on the device. Some error models require only one instrumentation function right before or after the targeted SASS instruction. Other error models are more challenging, since they require modifying an operand before the actual instruction execution and then restoring its content after the execution. These models are implemented with two instrumentation functions, plus a global memory storage mechanism to keep temporary data and communicate the functions during the runtime error propagation.

The software-level implementation details for each error model are described in the following, grouping the descriptions based on the technical similarities and highlighting their peculiarities. 


\footnotetext[1]{$M$ indicates a global memory location used for temporary data storage.}
\footnotetext[2]{The $[T_x, W_x]$ indicates the set of threads $T_x$ on selected warps $W_x$ where the error takes effect.}
\footnotetext[3]{$bitErrMask$ denotes the bits mask used to induce the index error to overwrite value.}
\footnotetext[4]{$Rd$ corresponds to the destination register.}
\footnotetext[5]{$R_{IR}$ represents the incorrect or invalid register to be accessed obtained from applying the $bitErrMask$ field to the original register number.}
\footnotetext[6]{$R<sx,sy,sz>$ denotes one of the source registers $R_sx$, $R_sy$, or $R_sz$.}

\textbf{IRA and IVRA}: Incorrect/Invalid Register Addressed Error models select a wrong register address in one of the operands fields for all instructions issued by the GPU. IRA selects an address that points to a valid wrong register (i.e., within the maximum number of registers per thread). IVRA selects registers outside of these boundaries as one of the operands. To implement IRA and IVRA, we use two different approaches; one is used when the corrupted register address represents the source operand and the other when the destination address is corrupted. 
The error descriptor for IRA and IVRA includes the parameters introduced in the section \ref{Code_instrumentation_methodology} (instruction, thread(s), warp(s) affected) plus additional parameters: \textit{bitErrMask}, and \textit{errOperloc}. The \textit{bitErrMask} is the bit level mask that modifies the target operand register number, and \textit{errOperLoc} is the operand position inside the instruction ($0$ means destination operand $Rd$, and $1$, $2$, or $3$ one of the source operands $R_{<sx,sy,sz>}$).

\begin{figure}
    \centering
    \footnotesize
    \begin{minipage}{0.47\textwidth}
    \begin{lstlisting}   
    (* $\cdots$ *) 
    /**Target SASS instruction*/
    (*\textbf{S2R} Rd, $SpecialRegisterID_{<x,y,z>}\footnotemark[7]$*)
    /** Error function */
    (*$Rd[T_x,W_x] \Leftarrow \ Rd[T_x,W_x]\footnotemark[2] \oplus bitErrMask[T_x,W_x]$*)
    (* $\cdots$ *) \end{lstlisting}
    \end{minipage}

\vspace{-4.5mm}
    
    \caption{Description of IAT/IAW/IAC error models.}
    \label{fig:IA-T-W-C}
    \par\vspace{-0.5\baselineskip}
\end{figure}
    \footnotetext[7]{$SpecialRegisterID_{<x,y,z>}$ refers to the special register SR\_TID or SR\_CTAID in any of the dimensions x, y or z}
    
    Fig.~\ref{fig:IRAerr} shows the implementation of the two operation modes of IRA/IVRA. The first mode refers to the error that targets the destination operands, thus the error function stores the content of the destination register $Rd$ into $M$ before the instruction is executed. Then, after launching the target instruction, the second instrumentation function copies into the target error register ($R_{IR}$) the result of the operation stored in $Rd$, then, the $Rd$ content is restored. 
    In the case of an error affecting the source operands, a function (issued before the instruction's execution) uses a memory location $M$ to store the content of the source register $R_{<sx, sy, sz>}$ before performing any data modifications. Then, the targeted register operand $R_{<sx, sy, sz>}$ takes the content of the error-accessed register $I_{IR}$. A second function, executed after the execution of the target instruction, restores the original source register $R_{<sx, sy, sz>}$ content. 


    
    \textbf{IAT, IAW, and IAC:} Incorrect Active Thread/Warp/CTA error models disable/enable or wrongly assign threads, warps, or CTA. To implement this behavior at the software level, we disable the execution of a set of threads on selected warp(s) by replacing their identifiers with different (wrong) ones, pointing threads to the same or different warps. 
  For example, for disabling thread0 in warp0, the index associated with the thread changes to the index of another thread (e.g., thread8 in warp0). Thus, the register that contains indexes for all threads will not contain the index of the disabled thread, producing the error effect during the execution by skipping the execution of thread0 in warp0. 

    Fig~\ref{fig:IA-T-W-C} presents the modeling concept of IAT, IAW, and IAC. This procedure is applied to the desired number of threads on selected target warp(s) $[T_x, W_x]$ issued on a specific SM sub-partition. It implements one instrumentation function after the instructions that copy the content of a special register $SpecialRegisterID_{<x,y,z>}$ into a destination register $Rd$. In the case of IAT or IAW, the instrumentation function affects only the instructions that take the content of SR\_TID for one of the x, y, or z dimensions of the parallel thread indexing of the application. The IAT (thread) error model keeps at least one thread active in the warp for its execution, whereas the IAW (warp) error model forces all indexes inside a warp to change, producing a full substitution of a particular warp for another. For IAC (CTA) error, the instrumentation function modifies the destination register $Rd$ of the instructions reading the SR\_CTAID special register of one of the thread's three dimensions x, y, or z indexing registers. In this case, when the index of the block changes, the obtained effect leads to incorrect block thread execution.

 
     \textbf{IAL:} The software-level implementation of Incorrect Active Lane error requires two different approaches. The first one, the unauthorized inactive lane (Fig.~\ref{fig:IALerr}.a), ignores the result of all instructions executed on a specific functional unit in one or several lanes (e.g., Integer or floating point cores). This functionality can be achieved by replacing the result of such instructions with the content of the destination register captured before executing the instructions. 
    The second approach (Fig.~\ref{fig:IALerr}.b) forces the execution of all predicated instructions associated with the Integer or Floating point Lane where the error is injected. An instrumentation function is inserted before the target instruction to check the predicate register status. Hence, if the predicate register \textit{disables} an instruction's execution, then    
    the function changes its status to \textit{enabled}, forcing the execution of an instruction that was not supposed to be executed.  
    
    \begin{figure}
    \centering
    \footnotesize 
    \begin{minipage}{0.47\textwidth}
    \begin{lstlisting}[title={a) Disable lane execution},abovecaptionskip={-2pt}]
    (* $\cdots$ *) 
    /** Error function: Part I*/
    (*$M \Leftarrow \ Rd[Lane,W_x]$*)
    /**Target SASS instruction*/
    (*\textbf{IMAD}*) Rd, Rx, Ry, Rz
    /** Error function: Part II*/
    (*$Rd[Lane,W_x] \Leftarrow \ M$*)
    (* $\cdots$ *) \end{lstlisting}
\end{minipage}\par\vspace{-1\baselineskip}

\begin{minipage}{0.47\textwidth}
    \begin{lstlisting}[title={b) Enable lane execution},abovecaptionskip={-2pt}]  
    (* $\cdots$ *) 
    /** Error function */
    (*\textbf{if} $Pr[Lane,W_x]== disabled$ \textbf{then}*)
        (*$Pr[Lane,W_x] \leftarrow enable$*)
    (*\textbf{end if}*)
    /**Target SASS instruction*/
    (* $<Pr>$ \textbf{IMAD}*) Rd, SrcOp_x, SrcOp_y, SrcOp_z
    (* $\cdots$ *) \end{lstlisting}
\end{minipage}

\vspace{-4.5mm}
    \caption{Description of IAL error models.}
    \label{fig:IALerr}
    \par\vspace{-1.5\baselineskip}
\end{figure}

\textbf{IIO, IMS, IMD, WV, and IOC:} All these errors modify a field in the executed instruction(s). 
These errors can be implemented by modifying the destination register of a selected group of instructions either with a random value or with a different operation, using the same instruction operands (see Fig.~\ref{fig:WVerr}). The instruction's group subject of the instrumentation and/or error injection is determined by the error type. Incorrect Immediate Operand (IIO) applies an error mask in the destination register for all instructions containing at least one reference to immediate operands. Incorrect Memory Source (IMS) inserts an error mask in all instructions containing at least one operand reference to constant or shared memory. Work-flow Violation (WV) selects and inserts an error mask in all instructions that write to a selected predicate register, affecting the application's control flow. Incorrect Memory Destination (IMD) targets all the instructions with shared memory as a destination reference by inserting an error bitErrMask either into the data register to be stored or in the register that addresses the shared memory. Finally, Incorrect Operation Code (IOC) targets all instructions issued by the integer or floating point cores by taking the input operands and replacing them with any other operation.

\begin{figure}
    \centering
    \footnotesize 
\begin{minipage}{0.47\textwidth}
  \begin{lstlisting} [title={a) IOC},abovecaptionskip={-2pt}]  
  (* $\cdots$ *) 
  /**Target SASS instruction*/
  (*\textbf{IADD}*) Rd, SrcOp_x, SrcOp_y
  /** Error function */
  (*$Rd[T_x,W_x] \Leftarrow SrcOp_{x[T_x,W_x]}$ ReplOp $SrcOp_{y[T_x,W_x]}$*)
  (* $\cdots$ *) \end{lstlisting}
\end{minipage}\par\vspace{-1\baselineskip}
\begin{minipage}{0.47\textwidth}
    \begin{lstlisting} [title={b) IIO/IMS},abovecaptionskip={-2pt}]  
    (* $\cdots$ *) 
    /**Target SASS instruction*/
    (*\textbf{IMAD}*) Rd, SrcOp_x, SrcOp_y, SrcOp_z
    /** Error function */
    (*$Rd[T_x,W_x] \Leftarrow Rd[T_x,W_x] \oplus bitErrMask[T_x,W_x]$*)
    (* $\cdots$ *) \end{lstlisting}
\end{minipage}\par\vspace{-1\baselineskip}

\begin{minipage}{0.47\textwidth}
    \begin{lstlisting} [title={c) IMD},abovecaptionskip={-2pt}]  
 (* $\cdots$ *) 
 /** Error function */
 (*$R_{<s,a>}[T_x,W_x] \Leftarrow R_{<s,a>}[T_x,W_x] \oplus bitErrMask[T_x,W_x]$*)
 /**Target SASS instruction*/
 (*\textbf{STS}*) [Ra], Rs
 (* $\cdots$ *) \end{lstlisting}
\end{minipage}\par\vspace{-1\baselineskip}
\begin{minipage}{0.47\textwidth}
    \begin{lstlisting}[title={d) WV},abovecaptionskip={-2pt}]      
    (* $\cdots$ *) 
    /**Target SASS instruction*/
    (*\textbf{ISETP}*) Pr, Rx, Ry, Rz
    /** Error function */
    (*$Pr[T_x,W_x] \Leftarrow \ Pr[T_x,W_x] \oplus bitErrMask[T_x,W_x]$*)
    (* $\cdots$ *) \end{lstlisting}
\end{minipage}
\vspace{-4.5mm}
    \caption{Description of IOC/IIO/IMS/IMD/WV error models.}
    \label{fig:WVerr}
    \par\vspace{-2.5\baselineskip}
   
\end{figure}

\begin{figure*}[ht]%
	\footnotesize
	\stackon[5pt]{}{\includegraphics[width=0.99\textwidth]{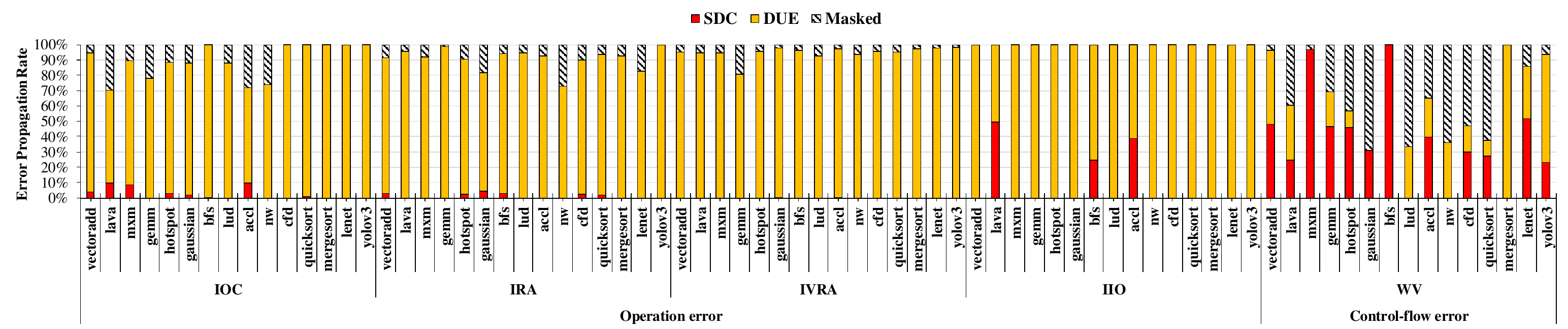}}%
	\\
	\stackon[5pt]{}{\includegraphics[width=0.99\textwidth]{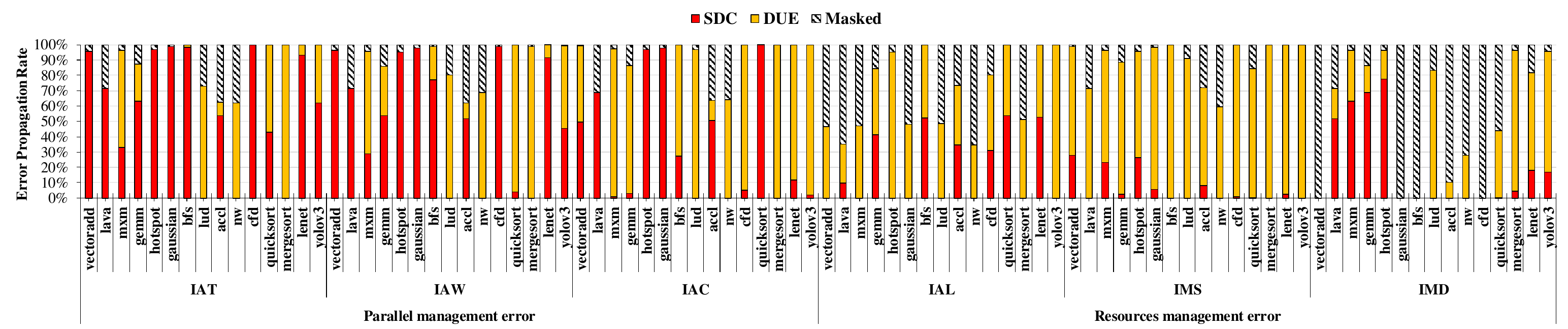}}%

\vspace{-2.5mm}

	\caption{{Error Propagation Rate} results of each error model propagated on 15 applications. IPP and IVOC are not shown since IPP is similar to other models and IVOC induces only DUEs.}
	\label{AVF_Software_Level}
    \par\vspace{-0.5\baselineskip}
    \end{figure*} 

    \textbf{IPP and IVOC:}
Incorrect Parallel Parameter (IPP) error has several ways of affecting the GPU operation, but most of them lay into two main categories \textit{i)} the wrong resource addressing hardware resources (i.e., registers or shared memory modeled by IRA, IVRA, IMS, and IMD), and \textit{ii)} by generating an incorrect threat execution modeled by IAT or IAW. On the contrary, Invalid Operation Code (IVOC) represents an invalid opcode operation that generates an invalid instruction exception at the software level, leading to a Device Unrecoverable Error in all cases the error is injected.
    
\subsection{Error propagation results} 
\label{sw_experimental_results}
    
    The software-based injection experiments have been performed on a workstation with an Intel i9-10900 CPU with 10  Cores, 32 GB of RAM, and one NVIDIA Ampere 3070ti GPU. We evaluated 15 real applications injecting 1,000  errors per application per error model. We target one sub-partition (PPB) of SM0. Overall, we inject more than 165,000 errors that took 300 hours of real GPU simulation. 

    
    Figure~\ref{AVF_Software_Level} reports, for the 15 applications, the \textit{Error Propagation Rate} (EPR), i.e., the probability for an error (produced by a fault that was activated and corrupted one or more of the unit outputs) to propagate to the software output. We plot the EPR for SDC, DUEs, and Masked. We group the results per error model. 
    As discussed in Section~\ref{low_experimental_results_Section}, we show the 11 error models grouped by the four main error groups (i.e., \textit{Operation Errors}, \textit{Control-flow Errors}, \textit{Parallel Management Errors}, and \textit{Resource Management Errors}). 
    We do not show IPP or IVOC error models since IPP can be implemented by any of the other error representations (IRA, IVRA, IAT, IAW, IMS, or IMD) and IVOC always generates DUEs at the low-level injections. 
    

    An interesting result from Figure~\ref{AVF_Software_Level} is the very high EPR for all error models and applications (the average EPR is 84.2\%). The applications that are either compute-intensive (i.g., yolov3, lava, or LeNet) or instance many kernels during the execution (i.g., bfs, mergesort,  and quicksort) present, for most of the error models, an EPR equal or close to 100\%. It is worth noting that permanent faults, by definition, are less likely to be masked compared to transient faults, as the resources are permanently damaged.   

    We can also see that the code's characteristics can significantly impact the EPR. This is particularly evident in two error models, WV (work-flow) and IMD (incorrect memory destination). For the WV error model, codes with many control flow blocks or thread indexing limitations that, once modified, can impact a significant amount of data (i.e., vectoradd, mxm, gemm, hotspot, bfs, and gaussian) show a high SDC EPR. Additionally, applications that can impact the memory addressing or block synchronization (i.e., lud, nw, and mergesort) show a high DUE EPR. The EPR is changed similarly for the IMD error model. For many codes, the error model IMD has no impact on the execution (i.e., vectoradd, gaussian, bfs, and cfd). The IMD error model affects instructions that operate on shared memories by changing the register that is the source or destination of an instruction that loads or stores on shared memory. Consequently, codes that do not use shared memories will have 100\% of the injected faults masked. 

\begin{figure}[t]%
        \centering
        \includegraphics[width=0.45\textwidth]{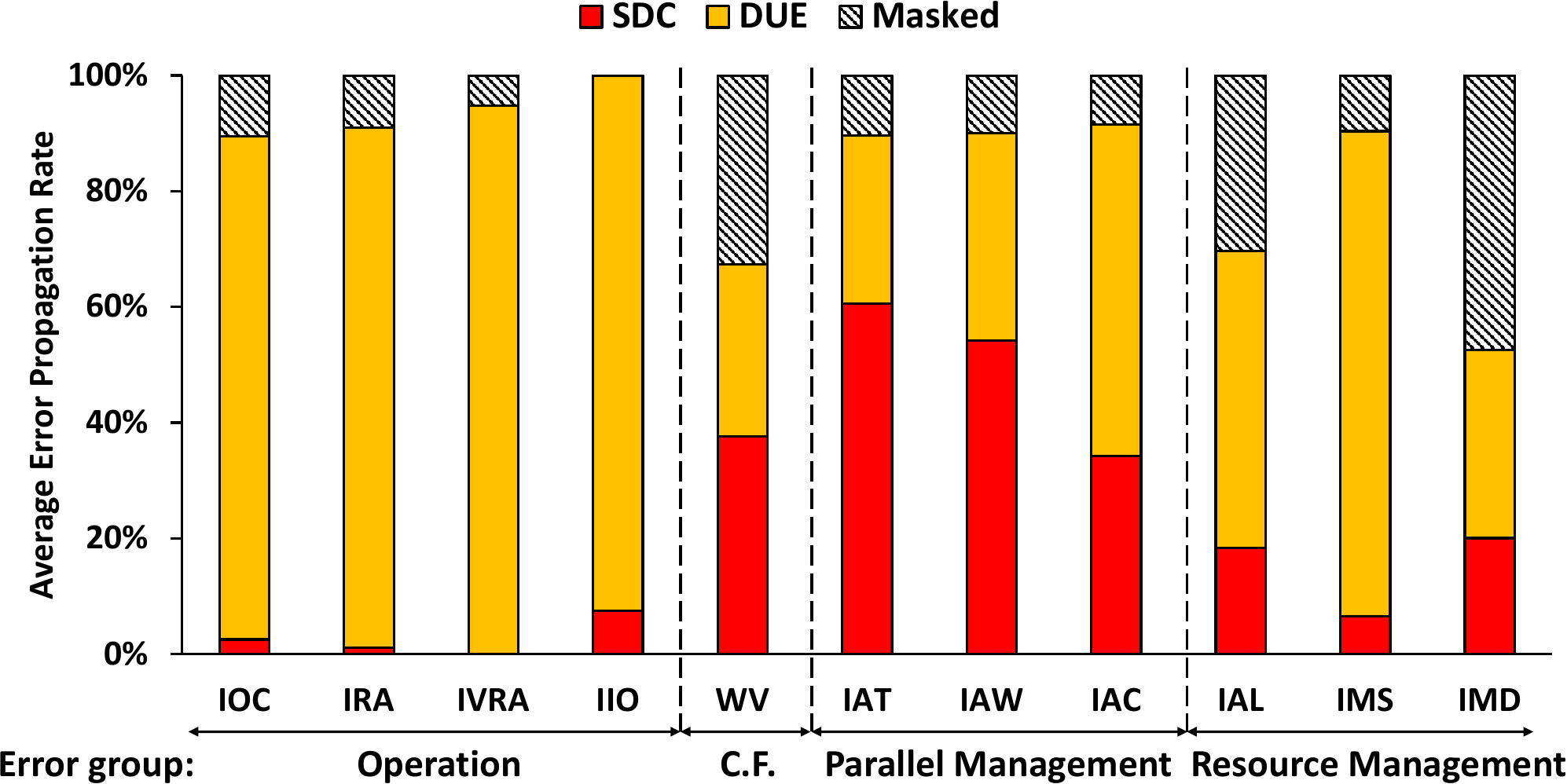}
        \caption{Average EPR among the 15 tested applications.}
        \label{figure_average_svf}
        \par\vspace{-1\baselineskip}
     \end{figure}

    Figure~\ref{figure_average_svf} summarizes the main findings for the 11 evaluated error models by showing the \textit{Average} EPR between all the codes.
    Interestingly, the group of \textbf{Operation Errors} shows a predominance of DUEs for all error models. On average, the percentage of IOC, IRA, IVRA, and IIO injections that generate a DUE is 87\%, 90\%, 95\%, and 92\%, respectively. The Operation Errors, as discussed in Section~\ref{error_model_implementation}, have a particular characteristic of modifying the behavior of all or many instructions in one or all threads within a warp or multiple warps. When many instructions are modified due to a permanent fault, the expected outcome is to have at least one thread, or many threads, performing illegal instructions, accessing incorrect memory accesses, or operating with registers outside the thread register bounds, which leads to a DUE. In fact, the percentage of incorrect memory addresses and illegal instructions generated by IOC, IRA, IVRA, and IIO error models are, on average, 99.05\%, 99.76\%, 100\%, and 98.29\% of the total DUEs.

    On the contrary, most of the error models that belong to the \textbf{Control-flow} and \textbf{Parallel Management} groups (WV, IAT, and IAW) have a high SDC EPR which is, on average, 38\%,  61\%, and 54\%, respectively. The combination of the error model and the executing code significantly changes these injections' outcomes. For example, when single or multiple threads are disabled due to the IAT error model, the output that would be expected from that thread will not be produced, generating an SDC. This is the case of codes like vectoradd, gaussian, cfd, and bfs, where the IAT error model enables/disables threads on the execution, and the code is able to finish (i.e., due to low interdependencies of the threads) but generates, most of the time, SDCs. Similar behavior is observed for IAW and WV error models, but in these cases, with a slightly higher incidence of DUEs than for IAT. In fact, these error models affect multiple threads or warps simultaneously, which can lead to the corruption of multiple output elements. 
    
    The errors from the \textbf{Parallel Management} group mostly induce SDCs in the applications. The only error model with an average DUE EPR higher than the SDC EPR is IAC (SDC EPR is 34\% and DUE EPR is 57\%). This happens because IAC is an error model that causes an incorrect execution (i.e., detention, assignation, or unauthorized submission) of an entire CTA (thread block) in the kernel execution, increasing the probability of DUEs. As with the other error models from the Parallel Management group, the EPR will change according to how the code uses the GPU resources. For instance, when the IAC error model is injected, applications such as lava, hotspot, gaussian, accl, and quicksort have most of the injections leading to SDC, i.e., the SDC EPR is 69\%, 97\%, 98\%, 51\%, and 99\% respectively. This happens because those applications schedule many independent parallel CTAs, then an incorrectly assigned block may still finish and produce an incorrect output.

    For the error models from the \textbf{Resource Management Errors} group, the EPR shows a strong dependence between the injection outcome and the code, where, for example, in the case of IMD, the use of shared memory can determine if the error will be masked or not. Similar behavior also can be observed for the IAL and IMS error models. However, as IAL and IMS affect resources used for all the codes (by disabling GPU lanes or causing an incorrect assignation of a memory resource for the result’s storing), we can see that both error models impact all codes, increasing their average EPR.

\subsection{Discussion}




Our method dramatically reduces the time complexity required for the evaluation of permanent faults in GPU, allowing an accurate error characterization at the gate level and a practical propagation of errors at the application level. A similar evaluation using only low-level hardware descriptions would be simply unfeasible. In fact, the simulation of a complete GPU at gate level takes, in our server, $\approx$14.5 hours for characterizing \textit{one} permanent fault in \textit{one} application. If we scale the simulation time for all workloads (15 applications) and all fault locations (50,044) we tested, we would reach a theoretical simulation time of around $14.5\times{}50,044\times{15}$ hours, that is $10.8\times{}10^6$ hours: $\approx$ 1,242 years!  In contrast, our approach required only 20.5 hours for profiling, 178.1 hours for low-level characterization, 4.2 hours for error analysis, and $\approx$ 300 hours of software-level error propagation for all workloads and targeted units (502.8 h in total), so speeding up the simulation of more than four orders of magnitude. 

The flexibility of our method allows its adaptation/extension for the evaluation of other units and fault models. The low-level micro-architecture characterization just requires the adaptation of the hardware profiling tool, according to the specifications of the new target unit or fault model, to identify the stimuli.



Correlating the low-level and software analyses we can identify the most probable instruction-level errors and the hardware units responsible for the observed application failures. From the low-level analyses (Table~\ref{tab:table_HW_fault_rate} and Figure~\ref{AVF_figure_micro_arch}), around 50\% of faults in \textit{fetch} and \textit{decoder} units produce visible instruction-level errors. These faults are mainly mapped as parallel management and operation error classes (from 1\% up to 15\% in the fetch, and from 12\% to 19\% in the decoder). Moreover, we observe that all propagated faults (30.5\%) in the WSC mainly affect the parallel management parameters (from 1\% to 9\%). Then, from the software level error injections (Figure~\ref{figure_average_svf}), we found that the parallel management and control flow errors are likely to induce SDCs (between 38\% to 60\%). The resource management errors produce mainly DUEs and 20\% of SDCs, while an operation error leads to DUEs in more than 90\% of the cases. 
Thus, permanent faults on the WSC are more likely to generate, at the application level, SDCs, whereas the permanent faults affecting the fetch unit lead, in more than 90\% of the cases, to DUEs (mainly due to illegal memory access). Finally, faults in the decoder unit have a higher probability of generating DUEs (70\%) and SDCs (20\%).

Our method can also support the design of detection and mitigation solutions for permanent faults. Adopting software detection techniques, in combination with smart scheduling and SMs swapping, allows fast fault detection or reduces the probability of permanent faults, thus potentially extending the in-field operation of GPUs. For example, we observed that most of the faults affecting the WSC generate SDCs in the application. Control-flow-checking strategies combined with smart thread scheduling replication can be a potential countermeasure against faults in the WSC. The control-flow-checking mechanism can be used to detect malfunctions during the execution of a given application. Then, the smart scheduling policy can discard the results from a faulty thread or warp by selecting the correct data from one of the replicated results. For the case of the Fetch and Decode units, since their corruption generates DUE, a hardware-base hardening technique is to be preferred. Given the Fetch and Decode units' importance and that a fault produces a catastrophic failure for all threads, it is unfeasible to use software-based mitigation strategies.

\section{Conclusions}
\label{sec_conclusion}


    We have proposed a method to understand permanent fault activation and propagation. 
    We exploit a multi-level approach to combine accurate gate-level simulations with efficient software-based error propagation. We focus on the WSC, Fetch and Decoder units and present the first quantitative estimation of their failure in the GPU code execution. We have identify four main groups of errors: (\textit{i)} Operation, \textit{ii)} Control-flow , \textit{iii)} Parallel management, and \textit{iv)} Resource management errors), corresponding to 13 instruction error categories. We have implemented instrumentation functions for the 13 error categories and used them to propagate the error effects in 15 real applications, resorting to a specially crafted instruction-level error injector (\toolname{}).

    
     The experimental results show that the permanent fault effect depends on the corrupted unit and executed instruction. Faults in the fetch unit mainly (66.80\%) lead to \textit{Operation} errors, faults in the Decoder unit lead to \textit{operation} (44.32\%) and \textit{resource management} (38.35\%) errors, and faults in the scheduler lead to \textit{parallel management} errors (54.87\%). 
    The software-level propagation of the observed error categories, shows that parallel management errors (mainly generated within the WSC) generate a high amount of SDCs (20\% to 60\%), whereas faults in the Fetch and Decoder  units mainly lead to DUEs ($>9$0\% and 70\%, respectively) . 

\bibliographystyle{ACM-Reference-Format}
\bibliography{references}






\end{document}